\documentstyle[a4wide,12pt,epsf]{article}

\parskip=\medskipamount
\newcommand{\be}{\begin{equation}}
\newcommand{\ee}{\end{equation}}
\begin{document}
\begin{center}
{\large
{\bf
Wigner distributions and the joint measurement of incompatible
observables
}}\\

\vspace{.3cm}

Willem M. de Muynck,\\
Department of Theoretical Physics,
Eindhoven University of Technology,\\
Eindhoven, The Netherlands
\end{center}

\begin{abstract}
A theory of joint nonideal measurement of incompatible observables is
used in order to assess the relative merits of quantum tomography and certain
measurements of generalized observables, with respect to completeness of
the obtained information. A method is studied for calculating a Wigner
distribution from the joint probability distribution obtained in a joint
measurement. 
\end{abstract}
\section{Quantum tomography}
The Wigner distribution $W(q,p)$ has the well-known Fourier
representation 
\[ W(q,p) 
 = \frac{1}{2 \pi^2}
\int_{-\infty}^{\infty} d\xi_1  \int_{-\infty}^{\infty}  d\xi_2
\tilde{W}(\xi)e^{i\sqrt{2}(p\xi_1 - q\xi_2)},\]
\[\tilde{W}(\xi) = Tr \hat{\rho} e^{(\xi \hat{a}^\dagger - \xi^*
\hat{a})},\; \xi = \xi_1 + i\xi_2 \]
(carets denote operators).
Putting $\xi = i\eta e^{i\theta}/\sqrt{2}$, it was observed by
Vogel and Risken \cite{VoRi} that 
$\tilde{W}(\xi)$ satisfies
\[ \tilde{W}(i\eta e^{i\theta}/\sqrt{2}) = Tr \hat{\rho} e^{i \eta
\hat{Q}(\theta)},\] 
which is the characteristic function of the rotated quadrature phase operator
\[ \hat{Q}(\theta)
= \frac{1}{\sqrt{2}} (\hat{a}^\dagger e^{i \theta} + \hat{a} e^{-i\theta}),
\]
measured in homodyne optical detection. 
Since the characteristic function is the Fourier transform of
the probability distribution, 
it was found \cite{VoRi} that the Wigner distribution can
be given as the integral
\[W(q,p) = \frac{1}{4\pi^{2}} \int_{-\infty}^{\infty}dx
\int_{0}^{2\pi} d\theta \int_{0}^{\infty} \eta d\eta 
e^{i\eta x- i\eta(p \sin \theta + q \cos \theta)}
w(x,\theta),\] 
in which $w(x,\theta)$ denotes the probability distribution of
the rotated quadrature phase observable $\hat{Q}(\theta)$.
The conclusion that can be drawn from this relation is that the
state of the system is completely determined if the probability
distributions of all the rotated quadrature phase operators are
known. Hence, these operators constitute a socalled quorum \cite{BaPa}.
It was remarked by Stenholm \cite{Stenholm92} that a
process of state determination along these lines is similar to
the one used in tomography, an obvious disadvantage being its
practical intractability because of the necessity of measuring the
full probability distribution $w(x,\theta)$ for {\em all} angles
$\theta$. 

\section{Generalized measurements}
It was also demonstrated 
by Stenholm \cite{Stenholm92} that an improvement in 
collecting information can be achieved by means of a {\em
simultaneous} measurement procedure of position and momentum
proposed by Arthurs and Kelly \cite{ArtKel}, the joint
probability distribution
$P(q,p)$ found in this measurement being expressible in terms of
the Wigner distribution $W(q,p)$ according to
\begin{equation}
P(q,p) = \frac{1}{2\pi^2}  \int_{-\infty}^{\infty} dQ \int_{-\infty}^{\infty} dP
e^{- \frac{(Q-q)^2}{s^2}} e^{-s^{2}(P-p)^2}
W(Q,P), 
\label{1}
\end{equation}
$s$ an arbitrary real parameter.
That the Arthurs-Kelly measurement procedure is indeed a {\em
complete} measurement, determining completely the state $\hat{\rho}$,
can be seen \cite{dMMa902} by inverting (\ref{1}) by means of deconvolution,
thus obtaining
\begin{equation}
W(Q,P) =
\frac{1}{2\pi} \int_{-\infty}^{\infty} dk
\int_{-\infty}^{\infty} dk' 
e^{\frac{k^2 s^2}{4}+ \frac{k'^2}{4s^2}}
\int_{-\infty}^{\infty} dq
\int_{-\infty}^{\infty} dp e^{ik(q-Q) + ik'(p-P)} P(q,p), 
\label{2}
\end{equation}
the $k$-integrals existing if the double Fourier transform of
$P(q,p)$ has asymptotic behaviour $o(\exp(-\frac{k^2 s^2}{4} -
\frac{k'^2}{4s^2}))$. It is not difficult to prove that this is
the case if $P(q,p)$ is given by (\ref{1}).

The probability distribution (\ref{1}) was
already found by Husimi \cite{Husimi}. Defining the squeezed
states $\phi^{s}_{q,p} (x)$ according to
\[
   \phi^{s}_{q,p} (x) =
   (\pi s^2)^{-1/4}
   e^{-\frac{(x-q)^2}{2s^2} + ix(p-q/2)},
\]
the Husimi distribution can be represented as
\be
  P(q,p) = Tr \hat{\rho}\frac{1}{2\pi} |\phi^{s}_{q,p}
  \rangle \langle \phi^{s}_{q,p}|.
\label{3}
\ee
Hence, the parameter $s$ in (\ref{1}) is just the squeezing
parameter. 

The essential point to be noted is, that a measurement yielding
the probability distribution (\ref{3}) is not described by a
projection-valued measure generated by the orthogonal spectral
resolution of some selfadjoint operator, but by a {\em positive
operator-valued measure} generated by the set of positive
operators
\be
\hat{M}(q,p) = \frac{1}{2\pi}
|\phi^{s}_{q,p} \rangle \langle \phi^{s}_{q,p}|,
\label{4}
\ee
satisfying
\[
\hat{M}(q,p) \geq \hat{O},\; \int^{\infty}_{-\infty}
dq \int^{\infty}_{-\infty} dp\; \hat{M}(q,p) = \hat{I}.
\]
Measurements described by positive operator-valued measures that
are not projection-valued measures are called {\em generalized}
measurements, measuring {\em generalized} observables. Such
measurements have been introduced by Davies \cite{Dav76}, Holevo
\cite{Hol82} and Ludwig \cite{Lud83}, and
are studied intensively by now. For instance, it has been
demonstrated \cite{WaCa,YuSha} that the eight-port homodyning
technique of detecting monochromatic optical signals, in which
the signal is mixed in a Mach-Zehnder interferometer with a
sufficiently strong local oscillator field of the same frequency
\begin{figure}[t]
\leavevmode
\centerline{
 \epsfysize=1.5in
 \epsffile{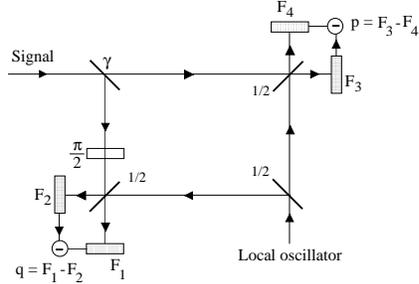} 
}
\caption{Eight-port optical homodyning as a joint nonideal measurement of
position and momentum observables}
\end{figure}
(cf. fig.1), yields (\ref{1}) for the joint probability of the
balanced signals $q = F_1-F_2$ and $p = F_3-F_4$, if the
transparencies of the mirrors are chosen as indicated in the
figure. The parameter $s$ is determined by the transparency
$\gamma$ according to
\[
s^2 = \frac{\gamma}{1-\gamma}.
\]

The technique of optical homodyning is known to induce excess
quantum noise in the signal. As a matter of fact, calculating
the marginals of $\hat{M}(q,p)$ we find
\[
\hat{M}(q) = \int^{\infty}_{-\infty}dp\;
\hat{M}(q,p) =
\int^\infty_{-\infty}dq'\; \frac{1}{\delta_1 \sqrt{\pi}} 
e^{-\frac{(q-q')^2}{\delta_1^2}}
|q' \rangle \langle q'|,\;\delta_1^2 = s^2,
\]
\[
\hat{M}(p) = \int^{\infty}_{-\infty}dq\;
\hat{M}(q,p) =
\int^{\infty}_{-\infty} dp'\; \frac{1}{\delta_2
\sqrt{\pi}} e^{-\frac{(p-p')^2}{\delta_2^2}}
|p' \rangle \langle p'|, \;\delta_2^2 = s^{-2}.
\]
This result can be interpreted as follows. If $s=0$ we have
$\hat{M}(q) = |q\rangle\langle q|$. Hence, if the partly
transparent mirror is replaced by a completely reflecting one
($\gamma = 0$), then the homodyning experiment can be interpreted as
an ideal measurement of the quadrature observable $\hat{Q} =
1/\sqrt{2} (\hat{a}^\dagger + \hat{a})$. On the other hand, this
measurement arrangement does not yield any
information on the canonically conjugated
observable $\hat{P} = i/\sqrt{2} (\hat{a}^\dagger - \hat{a})$.
If $\gamma = 1$, i.e., the partly transparent mirror is removed
completely, we have $s=\infty$, and the situation is now the
complementary one in which ideal information is obtained on
$\hat{P}$, whereas all information on $\hat{Q}$ is wiped out.
In the intermediate situations of finite $s$ we can interpret
the experiment as a joint nonideal measurement of $\hat{Q}$ and
$\hat{P}$, the Gaussian convolutions describing the excess noise
induced in the quadrature observables by inserting the partly
transparent mirror. Note that 
the uncertainties $\delta_1$ and $\delta_2$ satisfy the relation
\[
\delta_1 \delta_2 = 1, \]
thus exhibiting clearly the complementarity present in the joint
nonideal measurement of the two incompatible observables
$\hat{Q}$ and $\hat{P}$. It is important to note that the
complementarity that is involved here can be seen to have no
bearing on the initial preparation of the object, i.e., on the
state function $\hat{\rho}$, since it is a property of the generalized
{\em observable} (\ref{4}) alone. This is completely in accordance
with the interpretation of complementarity as a mutual
disturbance of measurement results in a joint measurement of
incompatible observables.

\section{Joint nonideal measurements of incompatible observables}
The notion of a joint nonideal measurement of two incompatible
observables was discussed in Martens and de Muynck
\cite{MadM90}. Restricting for simplicity to (generalized)
observables having discrete spectra, the observables represented
by the POVMs $\{\hat{Q}_m\}$ and $\{\hat{P}_n\}$ are said to be
jointly nonideally measurable if a bivariate POVM
$\{\hat{M}_{mn}\}$ exists, the marginals of which satisfying
\begin{equation}
\sum _n \hat{M}_{mn} = \sum_{m'}\lambda_{mm'}
\hat{Q}_{m'},\; \lambda_{mm'} \geq 0,\; \sum_m \lambda_{mm'} = 1,
 \label{4.1} \end{equation}
\begin{equation}
\sum_m \hat{M}_{mn} = \sum_{n'} \mu_{nn'}
\hat{P}_{n'},\; \mu_{nn'} \geq 0,\; \sum_{n} \mu_{nn'} = 1.
\label{4.2} \end{equation}
The nonideality matrices $(\lambda_{mm'})$ and $(\mu_{nn'})$
determine the nonideality of the measurements of observables
$\{\hat{Q}_m\}$ and $\{\hat{P}_n\}$, respectively. Often these
matrices are invertible, inverses satisfying 
\be
\sum_{m'} \lambda^{-1}_{m'm} = 1, \;
\sum_{n'} \mu^{-1}_{n'n} = 1, 
\label{5}
\ee
the matrix elements of the inverse matrices being, however, in
general not nonnegative. It is interesting to note that, if the
inverses both exist, then it is possible in principle to
calculate the probability distributions $\{Tr \hat{\rho}
\hat{Q}_m \}$ and $\{Tr \hat{\rho} \hat{P}_n \}$ from the
measured joint probability distribution $\{Tr \hat{\rho} \hat{M}_{mn}\}$,
thus obtaining from the joint {\em nonideal} measurement of
$\{\hat{Q}_m\}$ and $\{\hat{P}_n\}$ {\em exact} information on
their probability distributions. This, in principle, holds true
for the eight-port homodyning case, although in actuality the
inversion process may be hampered by incomplete knowledge of the
joint probability distribution $P(q,p)$.

Defining now an operator-valued measure according to 
\be
\hat{W}_{m'n'} = \sum_{mn} \lambda^{-1}_{m'n}
\mu^{-1}_{n'n} \hat{M}_{mn}, 
\label{6}
\ee
it can easily be verified that (\ref {6}) satisfies the
following relations:
\be
\sum_{n'} \hat{W}_{m'n'} = \hat{Q}_{m'}, 
\label{7}
\ee
\be
\sum_{m'} \hat{W}_{m'n'} = \hat{P}_{n'}, 
\label{8}
\ee
\be
\sum_{m'n'} \hat{W}_{m'n'} = \hat{I}.
\label{9}
\ee
Because of the possibility that some of the operators
$\hat{W}_{m'n'}$ are not positive, the operator-valued measure is
actually a quasi-measure. Since, because of (\ref{7}) through
(\ref{9}) its expectation values have all the properties of a
Wigner distribution, it was called a {\em Wigner measure}. Applying
this procedure to the eight-port homodyning POVM (\ref{4}) the
Wigner measure obtained in this way turns out to have the Wigner
distribution (\ref{2}) as its expectation value.

\section{Joint nonideal measurement of polarization observables}
As a further example we consider the joint nonideal measurement
of photon polarization observables. A nonpolarizing beam
splitter (transparency $\gamma$), either transmitting a photon
\begin{figure}[t]
\leavevmode
\centerline{
 \epsfysize=1.5in
 \epsffile{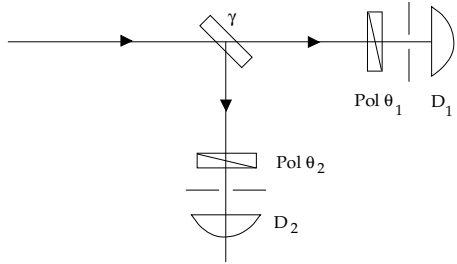} 
}
\caption{Joint nonideal measurement of two incompatible
polarization observables} 
\end{figure}
toward a polarizer having direction $\theta_1$ or reflecting it
toward a polarizer having direction $\theta_2$ (cf. fig. 2), can
be seen to realize a joint nonideal measurement of the
corresponding polarization observables. Denoting the spectral
representations of the two observables by $\{\hat{E}^1_-,\hat{E}^1_+\}$ and
$\{\hat{E}^2_-,\hat{E}^2_+\}$, respectively, the detection probabilities of
detectors $D_1$ and $D_2$ are given by 
$\gamma Tr \rho \hat{E}^1_m$ and $(1-\gamma) Tr \rho \hat{E}^2_n$, respectively, $m$ and $n$
both having the two possible values
``yes $= +$" and ``no $= -$" corresponding to the two
possible responses of the detectors. The joint detection probabilities for
the two detectors are then easily found as the expectation
values of the bivariate positive operator-valued measure generated by the
operators $\hat{M}_{mn}$ defined by
\be(\hat{M}_{mn}) =
\left( \begin{array}{cc}
\hat{O} & \gamma \hat{E}^1_{+}\\
(1-\gamma)\hat{E}^2_{+} & 1 - \gamma \hat{E}^1_{+} - (1 - \gamma) \hat{E}^2_{+}
\end{array} \right).\label{20} \ee
Calculating the marginals it is seen that (\ref{4.1}) and
(\ref{4.2}) are satisfied:
\begin{eqnarray}
\left( \begin{array}{c}
\sum_{n} \hat{M}_{+n}\\ \sum_{n} \hat{M} _{-n}
\end{array} \right) & = &
\left( \begin{array}{cc}
\gamma & 0 \\ 1-\gamma  & 1
\end{array} \right)
\left( \begin{array}{c}
\hat{E}^1_{+} \\ \hat{E}^1_{-}
\end{array} \right),
\label{(21)}\\
\left( \begin{array}{c}
\sum_{m}  \hat{M}_{m+}\\ \sum_{m}  \hat{M}_{m-}
\end{array} \right) & = &
\left( \begin{array}{cc}
1-\gamma & 0 \\ \gamma & 1
\end{array} \right)
\left( \begin{array}{c}
\hat{E}^2_{+} \\ \hat{E}^2_{-}
\end{array} \right).
\label{(22)}
\end{eqnarray}

From the inverses
\begin{eqnarray}
(\lambda^{-1}) =
\left( \begin{array}{cc}
\gamma^{-1} & 0 \\ 1-\gamma^{-1}  & 1
\end{array} \right), &
(\mu^{-1}) =
\left( \begin{array}{cc}
(1 - \gamma)^{-1} & 0 \\ 1 - (1-\gamma)^{-1}  & 1
\end{array} \right)
\nonumber \end{eqnarray}
of the nonideality matrices $(\lambda)$ and $(\mu)$ the Wigner
measure corresponding to this measurement arrangement 
can be found according to
\begin{equation}
(\hat{W}_{kl}) =
\left( \begin{array}{cc}
\hat{0} & \hat{E}^1_{+}\\
\hat{E}^2_{+} &  \hat{E}^1_{-} - \hat{E}^2_{+}
\end{array} \right).
\label{24.6}
\end{equation}
Note that, contrary to the eight-port homodyning case, the
expectation values of the
Wigner measure (\ref{24.6}) do not determine 
the state $\hat{\rho}$ completely. Hence, this measurement is
not a complete measurement.

It is not difficult to devise a polarization measurement that is
complete. Consider the arrangement of fig. 3 in which three
partly transparent mirrors are directing the photon toward one
out of four polarizers arranged along four different 
\begin{figure}[t]
\leavevmode
\centerline{
 \epsfysize=2in
 \epsffile{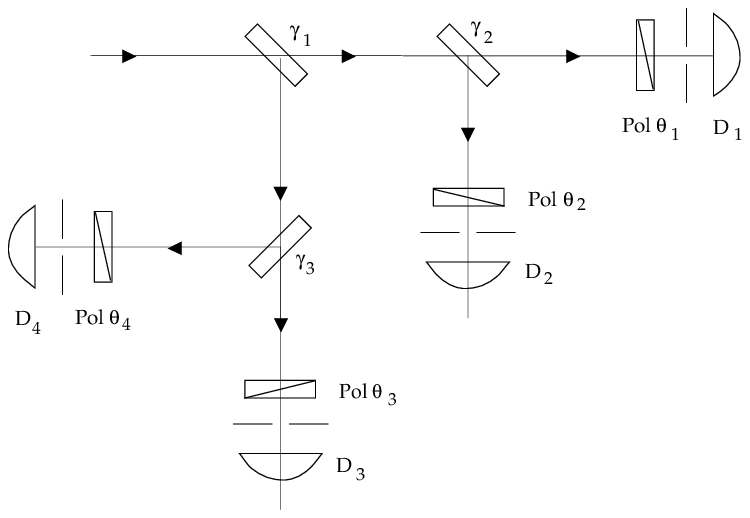} 
}
\caption{Joint nonideal measurement of four incompatible
polarization observables} 
\end{figure}
directions $\theta_1$
through $\theta_4$. The joint probability
distribution of this experiment can be found as the expectation
values of the operators
\begin {eqnarray}
\hat{M}_{+---} & = & \gamma_1 \gamma_2  \hat{E}^{1}_{+},\nonumber\\
\hat{M}_{-+--} & = & \gamma_1 (1 - \gamma_2)  \hat{E}^{2}_{+},\nonumber\\
\hat{M}_{--+-} & = & (1 - \gamma_1) \gamma_3  \hat{E}^{3}_{+},\nonumber\\
\hat{M}_{---+} & = & (1 - \gamma_1)(1 - \gamma_3)  \hat{E}^{4}_{+},\nonumber\\
\hat{M}_{----} & = & 1 - (\hat{M}_{+---} + \hat{M}_{-+--} +
\hat{M}_{--+-} + \hat{M}_{---+}),\nonumber 
\end{eqnarray}
generating the positive operator-valued measure describing the
experiment (all $\hat{M}_{ijk\ell}$ having more than one $+$ vanishing). 
We can order these operators in a way naturally generalizing (\ref{20}) so as
to obtain a {\em bivariate} POVM:
\begin{flushleft}$(\hat{M}_{mn}) =$\end{flushleft}
\footnotesize
\[\left( \begin{array}{cccc}
\hat{O} & \gamma_1 \gamma_2 \hat{E}^1_{+} &\hat{O}&\hat{O}\\
\gamma_1(1-\gamma_2)\hat{E}^2_{+} &\gamma_1[\hat{I} - \gamma_2 \hat{E}^1_{+} - (1 - \gamma_2)
\hat{E}^2_{+}]&\hat{O}&\hat{O}\\ 
\hat{O}&\hat{O}&\hat{O} & (1 - \gamma_1)\gamma_3 \hat{E}^3_{+} \\
\hat{O}&\hat{O}&(1 - \gamma_1)(1 - \gamma_3) \hat{E}^4_+ & (1 -
\gamma_1)[\hat{I} - \gamma_3 \hat{E}^3_{+} - (1 - \gamma_3) 
\hat{E}^4_{+}]
\end{array} \right).\]
\normalsize
Now this measurement can be seen to represent a joint nonideal measurement of
the observables represented by the POVMs 
\be\{\hat{Q}_m\} =
\{\gamma_1
\hat{E}^1_+, \gamma_1\hat{E}^1_-, (1 -\gamma_1)\hat{E}^3_+, (1 -
\gamma_1)\hat{E}^3_-\}\label{10}\ee
and
\be \{\hat{P}_n\} = 
\{\gamma_1 \hat{E}^2_+, \gamma_1\hat{E}^2_-, (1 -\gamma_1)\hat{E}^4_+, (1 -
\gamma_1)\hat{E}^4_-\},\label{11}\ee 
the two marginals $\{ \sum_n \hat{M}_{mn}\}$ and $ \{ \sum_m
\hat{M}_{mn}\}$ 
being expressible according to (\ref{4.1}) and (\ref{4.2}) in
the POVMs (\ref{10}) and (\ref{11}) with nonideality matrices
\begin{eqnarray*}
(\lambda) = \left( \begin{array}{cccc}
\gamma_2&0&0&0\\
1 - \gamma_2 & 1&0&0\\
0&0&\gamma_3&0\\
0&0&1 - \gamma_3&1
\end{array} \right),& (\mu) = \left( \begin{array}{cccc}
1 - \gamma_2&0&0&0\\
\gamma_2 & 1&0&0\\
0&0&1 - \gamma_3&0\\
0&0& \gamma_3&1
\end{array} \right).\end{eqnarray*}
Note that the POVMs (\ref{10}) and (\ref{11}) are maximal POVMs in the sense
that neither of them does describe a nonideal measurement of any inequivalent
generalized observable (cf. \cite{MadM90}).

The matrices $(\lambda)$ and $(\mu)$ are invertible. 
Calculating the Wigner measure (\ref{6}) for this measurement we find
\[ (\hat{W}_{mn}) = \left( \begin{array}{cccc}
\hat{O}&\gamma_1\hat{E}^1_+&\hat{O}&\hat{O}\\
\gamma_1\hat{E}^2_+&\gamma_1(\hat{E}^1_- - \hat{E}^2_+)&\hat{O}&\hat{O}\\
\hat{O}&\hat{O}&\hat{O}&(1 - \gamma_1)\hat{E}^3_+\\
\hat{O}&\hat{O}&(1 - \gamma_1)\hat{E}^4_+&(1 - \gamma_1)(\hat{E}^3_- -
\hat{E}^4_+) \end{array} \right).\]
It is easily verified that this Wigner measure has the 
POVMs (\ref{10}) and (\ref{11}) as its marginals. It is interesting to note
that, contrary to the joint nonideal measurement of two conventional
observables like those represented by the projection-valued measures
$\{\hat{E}^1_-,\hat{E}^1_+\}$ and 
$\{\hat{E}^2_-,\hat{E}^2_+\}$, the joint nonideal measurement of the two
{\em generalized} observables (\ref{10}) and (\ref{11}) is a complete measurement.

\end{document}